\documentclass[
]{ceurart}

\usepackage{listings}
\begin{document}

\copyrightyear{2023}
\copyrightclause{Copyright for this paper by its authors.
  Use permitted under Creative Commons License Attribution 4.0
  International (CC BY 4.0).}

\conference{Proceedings of the Work-in-Progress Papers at the 13th International Conference on Indoor Positioning and Indoor Navigation (IPIN-WiP 2023), September 25 - 28, 2023, Nuremberg, Germany}
\tnotetext[1]{Editors: Susanna Kaiser, Norbert Franke, and Christopher Mutschler}
\title{Ubiquitous Indoor Fine-Grained Positioning and Tracking: A Channel Response Perspective}


\author[1,2]{Chenglong Li}[
email=chenglong.li@nudt.edu.cn,
]
\cormark[1]
\address[1]{College of Electronic Science and Technology, National University of Defense Technology, 410073 Changsha, China}
\address[2]{WAVES group, Department of Information Technology, Ghent University-imec, 9052 Ghent, Belgium}

\author[2]{Emmeric Tanghe}[%
email=emmeric.tanghe@ugent.be,
]

\author[3]{Sofie Pollin}[%
email=sofie.pollin@kuleuven.be,
]
\address[3]{Wavecore, Department of Electrical Engineering, KU Leuven, 3001 Leuven, Belgium}

\author[2]{Wout Joseph}[%
email=wout.joseph@ugent.be,
]
\cortext[1]{Corresponding author.}

\begin{abstract}
  The future of location-aided applications is shaped by the ubiquity of Internet-of-Things devices. As an increasing amount of commercial off-the-shelf radio devices support channel response collection, it is possible to achieve fine-grained position estimation at a relatively low cost. In this paper, we focus on the channel response-based positioning and tracking for various applications. We first give an overview of the state of the art (SOTA) of channel response-enabled localization, which is further classified into two categories, \textit{i.e.}, device-based and contact-free schemes. A taxonomy for these complementary approaches is provided concerning the involved techniques. Then, we present a micro-benchmark of channel response-based direct positioning and tracking for both device-based and contact-free schemes. Finally, some practical issues for real-world applications and future research opportunities are pointed out.
\end{abstract}


\maketitle

\section{Introduction}
\begin{figure}[t]
\centering
\includegraphics[width=1\textwidth]{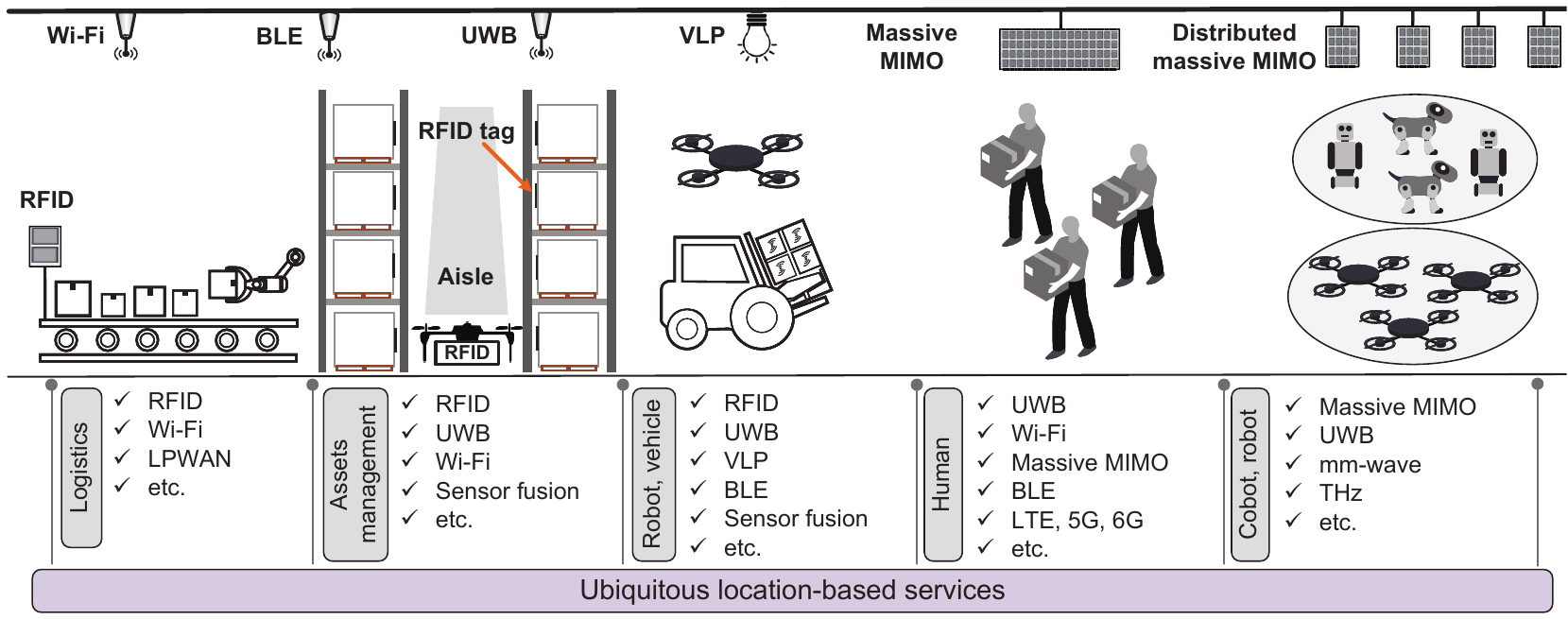}
\caption{Pervasive radio devices enabling various location-aided applications (RFID: radio frequency identification. LPWAN: low-power wide-area network. UWB: ultra-wideband. VLP: visible light positioning. BLE: Bluetooth low energy. LTE: long-term evolution. 5G: fifth-generation. 6G: sixth-generation. mm-wave: millimeter wave. THz: Terahertz.)}
\label{fig:PosTrackScenario}
\end{figure}

Location acquisition has obtained and continues to attract extensive attention due to the rapid proliferation of location-based services and the ubiquitous radio devices. Location awareness can help to automate and optimize the flows of many vertical applications \cite{Wollschlaeger2017,Chenglong2021Reloc2}, such as automatic inventory management, smart logistics, human-machine interaction, intelligent factory, etc., as shown in Fig.~\ref{fig:PosTrackScenario}. But for Global Navigation Satellite System-denied or indoor environments, accurate location acquisition is still challenging despite a range of radio frequency technologies and techniques that have been proposed and deployed during the past decades \cite{Zafari2019,You2021}. 
\par
Conventional localization solutions generally exploit the distance, angle, or Doppler information to estimate the location and trajectory. The localization accuracy highly depends on the corresponding ranging, angular, and velocity resolution. When relying on time-of-flight (ToF), the range resolution is directly related to the signal bandwidth as more bandwidth ensures finer-grained ToF and hence ranging. Angular resolution is increased when relying on a large antenna array with more antenna elements. However, it is not practical for bandwidth-limited systems, e.g., sub-6 GHz, due to the relatively large wavelength and thus large antenna sizes and inter-antenna spaces \cite{DeBast2022,Chenglong2022}. One potential solution to this problem is to construct the virtual antenna array of a moving antenna, namely built upon the idea of synthetic aperture\cite{Miesen2013}. Finer velocity resolution is also achieved with a higher frequency as Doppler is inversely proportional to the carrier wavelength. 
\par

Channel response has a finer granularity than the conventional signal strength-based metrics, because it can capture small-scale spatial motions via the complex signal (amplitude and phase) changes in the temporal, frequency, and antenna domains. For example, in the case of one-way propagation, half-wavelength distance changes will cause $\pi$-rad phase shifts. There is an increasing community, implementing localization systems based on channel response that is available in advanced communication systems \cite{Xiao2012, DeBast2022, Chenglong2022UWB, Altstidl2021, Miesen2013}. With the advances in radio hardware circuits and signal processing, many commodity radio devices support channel-sounding measurements, relying on channel state information (CSI) \cite{Xiao2012, DeBast2022}, channel impulse response (CIR) \cite{Chenglong2022UWB, Altstidl2021}, or signal phase \cite{Miesen2013}. These physical-layer measurements enable accurate characterization of signal propagation. Compared with the received signal strength indicator (RSSI), channel response provides a much finer-grained metric to describe the smaller scale spatial variations. It makes the human-environment and machine-environment interactions more feasible, enabling various emerging applications. 
\par
This paper is dedicated to positioning and tracking for the pervasive location-based services. According to the authors' best knowledge, there is limited work that overviews the state of the art (SOTA) from the perspective of the channel response. We classify the SOTA approaches into learning-based and model-based methods regarding device-based and contact-free positioning and tracking. A further taxonomy is presented based on how these leading-edge algorithms perform with the channel response. We conclude this paper by discussing some practical concerns and future research opportunities.

\section{Channel Response-based Indoor Positioning and Tracking}
\begin{figure}[t]
\centering
\includegraphics[width=0.8\textwidth]{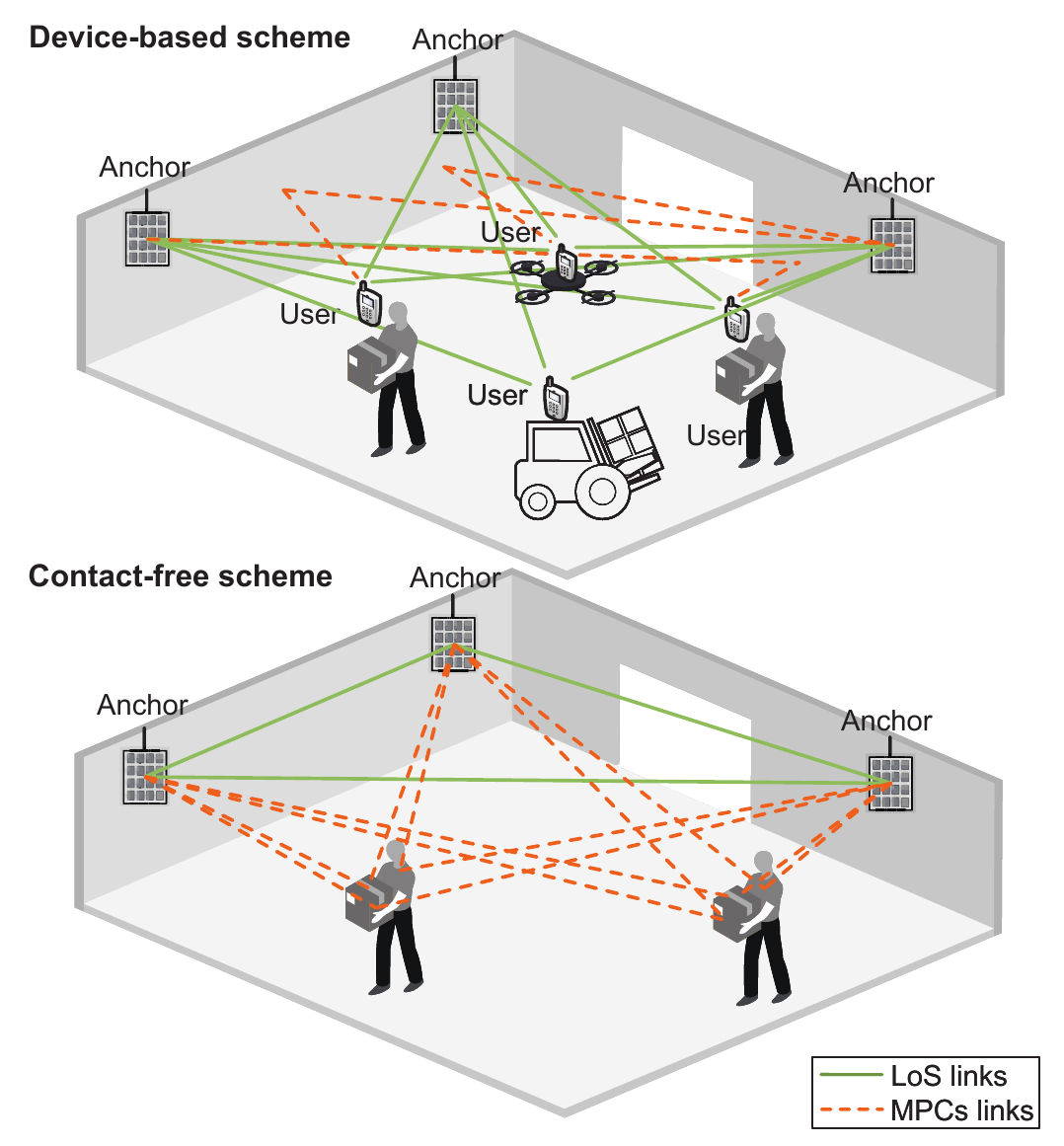}
\caption{Device-based and contact-free indoor positioning and tracking.}
\label{fig:PosTrackHuman3D}
\end{figure}

From the perspective of the radio device, indoor positioning and tracking can in general be classified into two categories, namely device-based and contact-free (\textit{a.k.a.} passive, device-free) schemes, as shown in Fig.\ref{fig:PosTrackHuman3D}. Device-based positioning and tracking is a cooperative scheme with radio sensors attached to the users or agents that should be tracked. With the help of channel-sounding data, device-based solutions have boosted the positioning accuracy from meter-level to decimetre-level despite limited bandwidth and the limited number of antennas. By contrast, contact-free positioning and tracking aim for the localization of non-cooperative users, which relies on radar-like sensors integrated in the environment to facilitate passive sensing. Again, due to the available channel-sounding data from the commodity devices, it is feasible to achieve positioning and tracking similarly as done for the phased array radar. For instance, we can do range-Doppler-angle processing on the channel response.
\par

Under assumption of a far-field MIMO channel, the channel transfer function at time $t$ and frequency $f$ (in equivalent base-band) can be given by
\begin{equation}\label{eq:channel}
{\bf{H}}(t,f)\!=\!\sum_{l=0}^L\!\alpha_l c_{\rm{Rx}}({\bf{\Omega}}_{{\rm{Rx}},l}) c_{\rm{Tx}}^{\top}({\bf{\Omega}}_{{\rm{Tx}},l})e^{\!-\mathcal{J}2\pi(\frac{d_l}{\lambda}+\nu_lt)},\!
\end{equation}where $\alpha_l, d_l, \nu_l$ denote the amplitude, distance, and Doppler shift of the $l$-th multi-path, respectively. $\lambda$ is the wavelength and $L$ the number of multi-path components. $c_{i}({\bf{\Omega}}_{i,l}), (i={\rm{Tx},\rm{Rx}})$ is the steering vector at transceiver and ${\bf{\Omega}}_{i,l}$ the directional vector. $l=0$ represents the line-of-sight (LoS) component. For device-based localization, if we can separate LoS from \eqref{eq:channel}, the location of the users can be inferred because ${\bf{\Omega}}_{{\rm{Rx}},0}$, ${\bf{\Omega}}_{{\rm{Tx}},0}$, $\tau_0$, and $\nu_0$ are the non-liner function of user's coordinates. 
\par
In the case of contact-free localization, the LoS component and the reflections or scatterings from the surroundings should be mitigated because the components associating with the target(s) are desired. Given several moving targets, the channel response of \eqref{eq:channel} can be rewritten as
\begin{equation}\label{eq:BS_CSI}
{\bf{H}}(t,f)\!=\!\bar{\bf{H}}(f)\!+\!\sum\limits_{j\in\mathcal{D}}\alpha_jc_{\rm{Rx}}({\bf{\Omega}}_{{\rm{Rx}},j}) c_{\rm{Tx}}^{\top}({\bf{\Omega}}_{{\rm{Tx}},j})e^{-\mathcal{J}2\pi\left(\frac{d_j}{\lambda}+\nu_jt\right)}\!,\!\!
\end{equation}where $\bar{\bf{H}}(f)$ is the sum of all static signals to be mitigated. $\mathcal{D}$ denotes the set of dynamic targets. To obtain the locations of the multiple targets, the ${\bf{\Omega}}_{{\rm{Rx}},j}$, ${\bf{\Omega}}_{{\rm{Tx}},j}$, $\tau_j$, and $\nu_j$ should be associated with the corresponding moving target and used for contact-free positioning and tracking.
\par
There are two popular technical routines for channel response-based localization. One is to estimate the distance, angle, or velocity-related metrics first, and then develop conventional positioning and tracking algorithms based on the estimated geometrical metrics. These distance- or angle-related metrics can be estimated from the CSI or CIR via super-resolution parameter estimation algorithms \cite{Kotaru2015,Chenglong2022}, such as multiple signal classification (MUSIC), space alternating generalized expectation-maximization (SAGE), a maximum likelihood parameter estimation framework (RiMAX), etc. The second is to exploit the channel response measurements directly for positioning and tracking purposes \cite{Garcia2017,Chenglong2023}, which evades the complex estimation of the intermediate metrics. The details of the taxonomy will be given in Section \ref{sec:SoTA}.

\section{Taxonomy of State-of-the-Art}
\label{sec:SoTA}
In this section, we will present the leading-edge channel response-based solutions regarding the device-based and contact-free schemes. The corresponding taxonomy of the channel response-enabling methods is shown in Fig.~\ref{fig:MethodCat} and will be further elaborated below.
\subsection{Device-based Positioning and Tracking}
\label{sec:DeviceBased}
For indoor applications, the device-based schemes rely on sensors or tags attached to the targets, like robots, (manned or unmanned) vehicles, packages, etc. As shown in Fig.~\ref{fig:PosTrackHuman3D} (top), the users infer their locations relative to the anchors from the received signal which includes the LoS and multi-path components (MPCs) from the surroundings. 

\subsubsection{Learning-based Approaches}
\begin{figure}[t]
\centering
\includegraphics[width=0.8\textwidth]{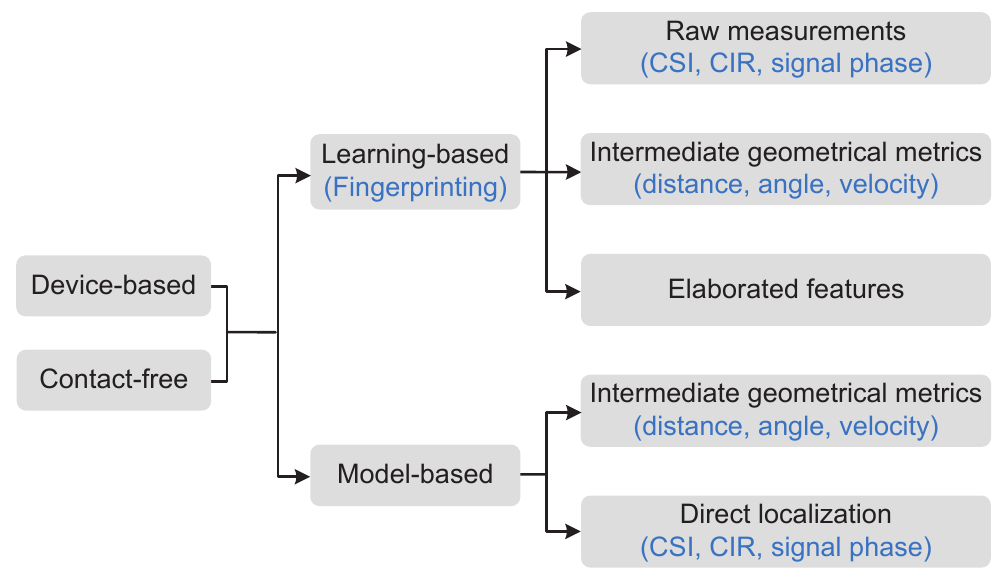}
\caption{Taxonomy of the channel response-based positioning and tracking.}
\label{fig:MethodCat}
\end{figure}

The learning-based (\textit{a.k.a.}, data-driven) scheme in Fig.~\ref{fig:MethodCat} generally establishes a non-linear mapping between the measurements and the targeted locations, which is intuitively easy to implement. Note that herein the fingerprinting localization approaches are also included within the learning-based framework because the off-line and on-line phases of fingerprinting are similar to the training and testing procedures, respectively. 
\par
\textbf{Raw measurements}: An intuitive positioning approach is to feed the raw channel measurements into a machine learning model directly to infer the coordinates of the users. Due to the pervasive access to Wi-Fi radios, there has been tremendous research focusing on CSI-based Wi-Fi localization in recent years. However, the measured channel response like CSI is a complex signal, and the widely used machine learning tools like neural networks generally are not capable of processing the complex values. Xiao \textit{et al}. \cite{Xiao2012} pioneered CSI-based fingerprinting localization system (FIFS) using merely the amplitude. Comparatively, Wang \textit{et al}. \cite{Wang2016} proposed a phase-based fingerprinting system, PhaseFi. A deep neural network with three hidden layers was adopted to train the phase data. Another possible solution is to feed the amplitude and phase (or the real and imaginary components) separately to the different channels of the convolutional neural network (CNN). Moreover, the measured CSI from the commodity devices is distorted due to the imperfect synchronization and other system errors, which requires complicated pre-processing. In \cite{DeBast2022}, De Bast \textit{et al}. showed the CSI can also be calibrated via deep learning based on an encoder-decoder architecture.
\par

\textbf{Intermediate geometrical metrics}: There are also many works on estimating the intermediate geometrical metrics, namely, angle, distance, or velocity, based on machine learning tools. In this case, machine learning acts as a super-resolution channel parameters estimation algorithm. After this, the estimated angle or distance metrics are exploited for positioning or tracking purpose. In \cite{Dai2022}, Dai \textit{et al}. proposed to use deep learning to derive angle-of-arrival (AoA) from a single snapshot of CSI, in which both classification- and regression-based models were investigated. Moreover, rather than estimating ToF from CIR directly, most of the learning-based research for ultra wide-band (UWB) localization focuses on the ToF or range correction \cite{Wymeersch2012}. 
\par
\textbf{Elaborated features}: Even though learning from raw channel measurements is straightforward to implement, it is not easy to learn the location-related features unless a deep neural network is adopted. It might be not feasible for lightweight mobile edge computing. In \cite{Chi2021}, Wu \textit{et al}. designed an angle-delay domain channel power matrix (ADCPM) for 3D massive MIMO localization. The ADCPM includes the multi-path power spectrum of delay, elevation, and azimuth in 4-D tensor, which achieved higher accuracy with reduced computational complexity and storage overhead. Another main drawback of learning-based approaches is the generalization ability for varied environments. To this end, another potential learning-based research direction is to elaborate the cross-domain features rather than using raw channel response.

\subsubsection{Model-based Approaches}
Learning-based solutions can achieve very high accuracy at the cost of labour-intensive data collection and labelling. Comparatively, the model-based approaches establish the explicit mathematical model from the channel response to the location, which are more physically interpretable. Furthermore, the model-based algorithms do not have the generalization problem as in the learning-based approaches, which are attractive from the perspective of real-world implementation.
\par
\textbf{Intermediate geometrical metrics}: Within this framework, the distance-, angle-, and velocity-related metrics are estimated from the channel response via the super-resolution parameters estimation algorithms, \textit{e.g.}, MUSIC, SAGE, RiMAX, etc. Then these intermediate geometrical metrics are used for positioning and tracking. Specifically, SpotFi was proposed based on commodity Wi-Fi devices \cite{Kotaru2015}. The AoA of the multi-path propagation was estimated from the CSI first, then the AoA of a direct link was obtained using a filtering method. Due to the high-ranging resolution of UWB, multi-path assisted localization has attracted attention by exploiting the ToF estimates of both LoS and the specular multi-path reflections. Generally, given the prior knowledge of the floor plan or within the simultaneous localization and mapping framework \cite{Witrisal2016}, the fixed but unknown virtual anchors' locations can be obtained. Such this, positioning accuracy can be enhanced via the increasing spatial diversity of more (virtual) anchors.
\par
\textbf{Direct localization}: Another promising solution is based on the channel response directly, as it depicts the rich propagation information and is also tightly location-related. This idea is especially suitable for bandwidth-limited systems, like ultra-high frequency RFID systems, sub-6 GHz Wi-Fi, BLE, etc. A promising accuracy can be achieved by exploiting the idea of synthetic aperture, namely the distributed antenna array or moving antenna. For example, in \cite{Miesen2013}, Miesen \textit{et al}. proposed to localize the RFID tag via a moving antenna on a robot arm. The moving antenna constructed a virtual antenna array, and the RFID tag was localized via a matched filter.

\subsection{Contact-Free Positioning and Tracking}
Besides the device-based localization mentioned above, contact-free localization also plays an important role in our daily life. Contact-free localization exploits the signal reflected or scattered by the targets to estimate the locations, such as worker safety monitoring and accident alerts in the assembly area in the industry, as shown in Fig.~\ref{fig:PosTrackHuman3D} (bottom). This is more feasible as it does not require attaching any radio devices or sensors to the workers. Meanwhile, it also preserves personal privacy without the device-based individual identity. 
\par
While conventional contact-free methods primarily process the signal strength variations for coarse positioning, the channel response in commodity devices enables small-scale motion sensing like the prohibitively large phased array radar with a relatively low cost. Besides wide-band radio systems like mm-wave devices, narrow-band radio systems are also able to achieve accurate spatial sensing when deploying multiple antennas, or a motorized platform based on the idea of synthetic aperture radar. As shown in Fig.~\ref{fig:MethodCat}, there also are learning-based and model-based approaches generally.
\subsubsection{Learning-based Approaches}
Contact-free positioning is generally more complicated than the device-based scheme as it utilizes the reflection or scattering from the targets. The target-of-interest signal is hidden in the complex multi-path signal and has lower power than the LoS components and even the MPCs from the surroundings. Learning-based approaches can mitigate or relieve the interference via a data-driven model.
\par
\textbf{Raw measurements}: Similar to the device-based scheme, learning from raw channel measurements for contact-free localization is relatively intuitive. The difference is the learning-based contract-free methods should mitigate the undesired background signals including the LoS propagation and the scattering from the surroundings. There are also mainly two categories of learning-based approaches: classification and regression. Classification-based methods are to compare the instantaneous measurements with the fingerprinting database or radio maps to infer the target's location. Regression-based methods are to infer the map from the raw channel measurements to the target's coordinates.
\par
\textbf{Intermediate geometrical metrics}: Machine or deep learning tools are exploited to estimate the distance, angle, or velocity metrics based on the channel measurements in the cases of absent and present targets. In \cite{Chenglong2022UWB}, Li \textit{et al}. proposed a residual CNN-based reflected ToF estimation method via learning the differences between static and dynamic CIR or corresponding variance sequences. Then a particle filter was implemented for passive human tracking based on the ToF estimates. Moreover, it concluded that the variance-based metric is less domain-dependent than using the CIR directly.
\par
\textbf{Elaborated features}: To advance cross-domain sensing and achieve (near) zero-effort sensing for new environments and system settings, a domain-independent feature, namely, the so-called \textit{body-coordinate velocity profile}, was designed for the gesture classification \cite{Widar3}. However, according to the authors' best knowledge, there is no available literature on contact-free positioning and tracking exploiting the elaborated domain-independent features. But it can be a promising direction concerning the scalability of the learning-based solutions.
\subsubsection{Model-based Approaches}
The taxonomy of model-based contact-free positioning and tracking is given as follows.
\par
\textbf{Intermediate geometrical metrics}: One of the most popular approaches is to estimate the reflected angle, distance, or Doppler velocity from the target(s) using the super-resolution parameters estimation algorithms. In \cite{Qian2018}, Widar 2.0 exploited the conjugate multiplication between the adjacent antennas to mitigate the CSI measurement errors and static interference, and applied the SAGE algorithm to estimate the targeted MPCs. A binary optimization, together with a Hungarian algorithm, was used to associate the targeted MPCs along the trajectory. Instead of using CSI multiplication, Wu et al. \cite{Wu2021} proposed a CSI-quotient model to estimate the targeted Doppler shift. The CSI-quotient model assumed the adjacent antennas possess the same static interference, which can obtain a higher targeted signal-to-noise ratio.
\par
\textbf{Direct localization}: By using the intermediate geometrical metrics or the MPCs estimates for contact-free tracking, we not only obtain the targeted MPCs but also the cluttered components. It is necessary to associate the targeted MPCs along the time, namely evolving association. If there are multiple targets, associating the MPCs with the specific target, the so-called target association, is also required. Furthermore, if the radar-like sensors are distributed in the environment, we estimate the MPCs for each sensor. In this case, we also need to associate the MPCs of a specific target with all distributed sensors, that is sensor or link association. When taking this three-dimension association problem into account, the contact-free positioning and tracking will be computing complex and difficult to resolve \cite{Chenglong2023}. Recently, there is work focusing on direct channel response-based contact-free localization without the consideration of MPCs association. Sakhnini \textit{et al}. \cite{Sakhnini2022} established a radar-like tracking prototype based on a massive MIMO communication test bed. However, a single metallic cylinder target was considered and localized.

\section{Micro-Benchmark: Channel Response-based Direct Positioning and Tracking}
Channel response-based positioning and tracking definitely is a promising research field considering the pervasive applications, especially, when an increasing number of commodity radio devices provide the access to channel measurements. Among them, channel response-based direct localization obtains the location estimates through the integration of physical-layer channel information, which show its priority concerning low signal-to-noise ratio and multi-path effect \cite{Garcia2017,Chenglong2023}. For device-based localization, instead of estimating the intermediate angle and distance metrics, we exploit the geometrical diversity of the multiple antenna. The user or target is localized via a matched filter built upon the idea of synthetic aperture. The user's location $\mathbf{P}_{\rm{U}}$ can be estimated via
\begin{equation}\label{eq:Pos_est}
\begin{aligned}
\!\!{\hat{\mathbf{P}}_{\rm{U}}}=&\mathop {\arg\max}\limits_{\mathbf{P}_{\rm{U}}}\mathcal{C}(\mathbf{P}_{\rm{U}})\\
=&\mathop {\arg\max}\limits_{\mathbf{P}_{\rm{U}}}\left\vert\sum\limits_{i=1}^{N_{\rm{Tx}}}\sum\limits_{j=1}^{N_{\rm{Rx}}}\sum\limits_{k=1}^{N_{\rm{f}}}\tilde{\alpha}_{i,j,k}e^{\mathcal{J}\big(\phi_{i,j,k}(\mathbf{P}_{\rm{UE}})-\tilde{\phi}_{i,j,k}\big)}\right\vert,\!\!
\end{aligned}
\end{equation}where $\tilde{\alpha}_{(\cdot)}$, $\tilde{\phi}_{(\cdot)}$ denote the amplitude and the phase of the measured channel response. $N_{\rm{Tx}}$, $N_{\rm{Rx}}$, and ${N_{\rm{f}}}$ are the number of transceiver's antennas and sub-carriers, respectively. $\phi_{\cdot,\cdot,\cdot}(\mathbf{P}_{\rm{UE}})$ is the phase calculated via the location candidates.
\par
For contact-free localization, we can adopt a similar matched filter as in \eqref{eq:Pos_est}, but instead the $\tilde{\alpha}_{(\cdot)}$, $\tilde{\phi}_{(\cdot)}$ denote the amplitude and the phase of the targeted channel response after interference mitigation. Define the range-Doppler profile at a single antenna as $\hat{\mathbf{H}}_{(i,j)}=\mathbb{T}^{\rm{H}}{\mathbf{H}}_{(i,j,\cdot,\cdot)}\mathbb{D}$, where $\mathbb{T}$ and $\mathbb{D}$ denote the discrete Fourier transform matrix, then we can obtain the targeted channel response $[\hat{\mathbf{H}}_{\rm{Target}}]_{(i,j)}\!=\!\hat{\mathbf{H}}_{(i,j,\cdot,\cdot)}(\hat{n}_{\rm{T}},\hat{n}_{\rm{D}})$, where
\begin{equation}\label{eq:Range_DFS}
\!\!(\hat{n}_{\rm{T}},\hat{n}_{\rm{D}})\!=\!\mathop {\arg\max}\limits_{n_{\rm{T}},n_{\rm{D}}}\mathbb{E}\left\{(\mathbb{T}^{\rm{H}}{\mathbf{H}}_{(i,j,\cdot,\cdot)}\mathbb{D})\odot(\mathbb{T}^{\rm{H}}{\mathbf{H}}_{(i,j,\cdot,\cdot)}\mathbb{D})^{*}\right\},\!\!
\end{equation}and $\odot$ denotes the Hadamard product. Note that the maximization in \eqref{eq:Range_DFS} indicates only single-target tracking is applicable through this method. For multi-target contact-free tracking, the targets association is needed. To handle this problem, compressive sensing and random finite set theory can be implemented, which is omitted here but refer to our latest work in \cite{Chenglong2023} due to page limitation. 
\par
To track the user or target, we propose to track the changes of the matched filter in \eqref{eq:Pos_est} through particle filter. For each location update, the $K$ particles are weighted via the differences with the maximum of $\mathcal{C}(\mathbf{P}^{(i)}_{\rm{U}})$, given as \cite{Chenglong2023},
\begin{equation}\label{PF_weight}
\begin{aligned}
\hat{P}_w^{(i)}=&\exp{\left(\frac{1}{2\sigma_c^2}\left(\min\limits_{1\leq i\leq K}\left\lbrace{P_w^{(i)}}\right\rbrace-P_w^{(i)}\right)\right)},\\
{\rm{s.t.}} \qquad\qquad &P_w^{(i)}=\left(\mathcal{C}_{\rm{max}}-\mathcal{C}(\mathbf{P}^{(i)}_{\rm{U}})\right)^2,
\end{aligned}
\end{equation}where $\sigma_c$ is the standard deviation of $\mathcal{C}(\mathbf{P}_{\rm{U}})$. $\mathcal{C}_{\rm{max}}\!=\!1$ because $\mathcal{C}(\mathbf{P}_{\rm{U}})$ is normalized to $(0,1]$.
\par
\begin{table}[t]
\centering
\caption{Parameters settings of the massive MIMO-OFDM system}
\begin{tabular}{l|c}
\hline
\textbf{System parameters} & \textbf{Value}          \\ \hline\hline
\textbf{Transmitter power} & 15 dBm					\\ \hline
\textbf{Sampling rate}     & 100 Hz  				\\ \hline
\textbf{Wavelength}        & 11.49 cm				\\ \hline
\textbf{\# of Sub-carrier}  & 100     				\\ \hline
\textbf{\# of antenna}     & 64 	  				    \\ \hline
\textbf{Modulation}        & QPSK    				\\ \hline
\textbf{Patch antenna size}& 7 cm$\times$7 cm  		\\ \hline
\textbf{Size of experimental space}& 6.5 m$\times$10 m  \\ \hline
\end{tabular}
\label{table:Paras}
\end{table}

\begin{figure}[t]
\centering
\includegraphics[width=0.99\textwidth]{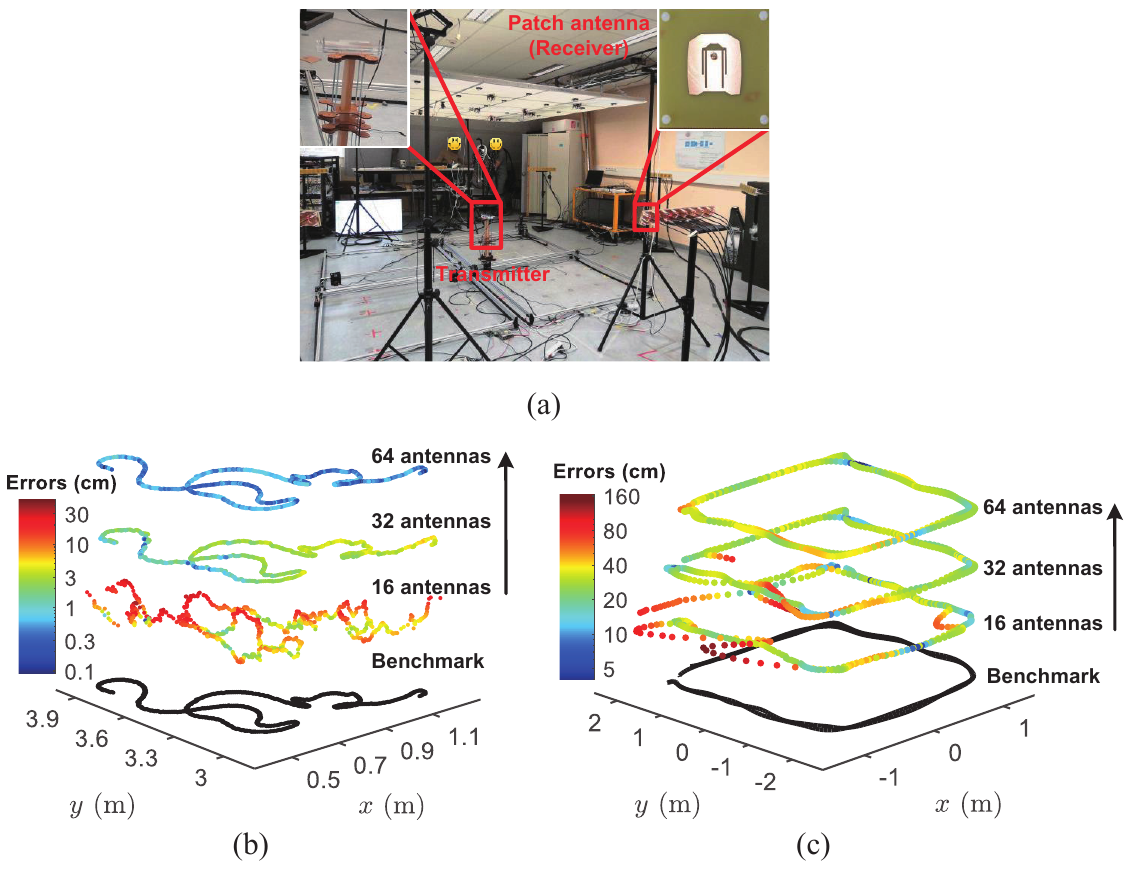}
\caption{Sub-6~GHz massive MIMO-OFDM system for positioning and tracking: (a) Measurement setup. (b) Device-based scheme. (c) Contact-free scheme.}
\label{fig:MaMIMO}
\end{figure}

 Fig.~\ref{fig:MaMIMO} shows the device-based and contact-free positioning and tracking results for a single user or pedestrian based on a sub-6~GHz distributed massive MIMO orthogonal frequency-division multiplexing (OFDM) communication test bed. The center frequency is 2.61~GHz and the bandwidth 18~MHz. The other parameter settings are given in Table \ref{table:Paras}. The transmitter is with a dipole antenna and the receiver with 64 patch antennas. 
\par
For both device-based and contact-free schemes, the impact of the number of antennas on accuracy is evaluated. The tracking accuracy is represented by the absolute distance errors between the tracking results and the ground truth. For the device-based scheme, the ground truth was obtained via the computerized numerical control X-Y table which guarantees a millimetre-level accuracy of the user equipment. For the contact-free scheme, the ground truth of the pedestrian was inferred via the active tracking. For the active tracking, we asked each participant to place the antenna (another transmitter) on the top of the head to avoid the possible body shadowing effect. The active tracking ensures a centimetre-level accuracy that has been testified in \cite{Chenglong2023}, which is sufficient and applicable for pedestrian tracking given the size of human body. As expected, the accuracy increases as more antennas have been exploited as shown in Figs.~\ref{fig:MaMIMO}(b)-(c). Note that we have not considered the antenna deployment optimization herein but based on the sequential antenna selection. For device-based tracking, when 64 antennas are adopted, a (sub-) centimetre-level accuracy can be achieved. Fig.~\ref{fig:MaMIMO}(c) shows the results of contact-free pedestrian tracking, in which a decimetre-level accuracy can be guaranteed under the assumption of cylindrical body model.
\section{Discussions and Future Opportunities}
Channel response makes the finer-grained positioning and tracking possible and applicable. However, there are still many challenges for real-world implementation. In the following subsections, we will discuss some open challenges and future research opportunities for channel response-based positioning and tracking.
\subsection{Probabilistic Positioning and Tracking}
Most available localization algorithms are based on such as the AoA and ToF estimates. However, these distance and angle metrics are only intermediate estimates from which the position is derived. Their estimation accuracy may be constrained by the resolution of the estimation algorithms. Meanwhile, the super-resolution algorithms generally require complex computations. One recently emerging solution for accuracy enhancement is based on soft information \cite{Conti2021} instead of using a single-value distance or angle estimate. The ranging or angular likelihood is extracted from the channel response measurements. The likelihood is generally assumed as a Gaussian mixture model, which can be solved via the expectation-maximization algorithm or learned by deep probabilistic learning. On the other hand, soft-ranging or soft-angular information rarely appears in contact-free positioning and tracking, which can also be a promising research direction. 
\subsection{Simultaneous Localization and Mapping}
Simultaneous localization and mapping is to construct the 2-D or 3-D map of an unknown environment and localize and track the targets simultaneously. To sense the surroundings, it requires a system that has a very high spatial resolution, such as UWB and mm-wave systems. With the environment knowledge, we can exploit the specular multi-path to enhance the localization performance with fewer anchors, so less human exposure as well \cite{Witrisal2016}. However, obtaining the information on the complete surrounding is not necessary for positioning and tracking. Abundant knowledge will also increase the computing complexity and the inference latency. For example, if we use the specular MPCs to help localization, we only need the location of the planar reflectors (e.g., wall). On the other hand, MPCs exploitation requires complex association. The factor graph model and random finite set filtering are the two promising techniques for this issue.
\subsection{Contact-Free Multi-Target Positioning and Tracking}
Contact-free localization becomes a hot topic due to the advance in integrated communication and sensing, which achieves the sensing functionality by reusing the allocated spectrum, hardware, and even signalling resources of the communication systems. First, background removal or ToI extraction is a big challenge for contact-free localization. The CSI-quotient model \cite{Wu2021} can effectively mitigate the undesired signal using the CSI ratio of adjacent antennas. Second, according to the authors' best knowledge, most work on contact-free localization, especially for sub-6 GHz systems, concentrates on a single target. How to generalize to an unknown number of targets or multiple targets desires further investigation. The widely adopted MPCs-based localization requires complex MPCs association for multiple targets. The problem becomes more difficult if distributed sensors are adopted. Third, investigations on contact-free multiple static targets or hybrid static-dynamic targets localization are rare. It is challenging to use the Doppler shifts or channel variations for interference mitigation. One possible solution is to use high spatial resolution radio to generate the points cloud or image of the targets. Another method is to detect the vital sign (vibration) of the human (machine). However, how to evade the interference from the unexpected motion or dynamic targets needs further exploration.

\subsection{Networked Positioning and Tracking}
When in a cluttered environment, the signal is easy to be blocked by the metallic assets, which degrades the localizability and the positioning accuracy. On the other hand, for the sub-6 GHz bandwidth-limited systems, the angular resolution is limited even though a small antenna array can be installed. Therefore, improving the geometrical diversity of the anchors or access points can enhance the accuracy and coverage. During tracking, the local access points can share the channel information with the remote access points as a priori knowledge. Furthermore, for multi-target device-based tracking, the localized targets can also act as the new access points if they possess the transceiver functionality, namely, the networked positioning and tracking. Through information exchange, the field of sensing and the performance of each access point can be improved greatly.

\subsection{Domain-Independent Positioning and Tracking}
Even though the learning-based approaches are straightforward to implement and fairly easy to use, one of the main drawbacks of this scheme is that generalization for a new system setting (including hardware heterogeneity, antenna orientation, etc.) or a new environment is challenging. They are namely \textit{domain dependent}. Training from scratch in a new domain or involving transfer learning can solve this problem, but it requires newly labelled dataset collection which adds effort more or less to practical implementations. To handle this problem, one solution is to establish an open-access dataset that is contributed by the whole community, like the well-known image recognition dataset ImageNet (https://www.image-net.org/). The open-access dataset should include the measurements at different sites, using different commodity devices, having different numbers of anchors, etc. The dataset provides a cross-domain platform for algorithm development and evaluation. It can also help to mitigate the performance gap from the well-controlled laboratory to practical implementation. Another potential learning-based research direction is to feed the less domain-dependent metrics to the model instead of the raw channel response. The domain-independent metrics generally require elaborated design, which should retain the rich channel information but be less dependent on the system setting or environment.

\section{Conclusion}
Location acquisition has become an important content considering the ubiquitous location-aided vertical applications. Physical-layer channel information, i.e., channel response, in commodity devices allows us to better understand the signal propagation and develop the algorithms for positioning and tracking purposes. In this paper, the representative SOTA regarding device-based and contact-free schemes are classified into learning-based and model-based approaches. The corresponding pros and cons are discussed and analysed. This paper also presents of channel response-based direct positioning and tracking, which provides a micro-benchmark for both device-based and contact-free schemes. Furthermore, we will point out some limitations that the available work that have not solved well for channel response-enabled localization. Some open challenges and future opportunities are given as well.

\section*{Acknowledgments}
This work is supported in part by the Excellence of Science (EOS) project MUlti-SErvice WIreless NETworks (MUSE-WINET), in part by the imec.icon project InWareDrones, in part by the imec project UWB-IR, and in part by the Research Foundation Flanders (FWO) under Grant no. G098020N. The authors would like to thank Sibren De Bast and Robbert Beerten from KU Leuven for the help of the distributed massive MIMO experiments.

\bibliography{PosTrack}

\begin{thebibliography}{23}
\expandafter\ifx\csname natexlab\endcsname\relax\def\natexlab#1{#1}\fi
\providecommand{\url}[1]{\texttt{#1}}
\providecommand{\href}[2]{#2}
\providecommand{\path}[1]{#1}
\providecommand{\DOIprefix}{doi:}
\providecommand{\ArXivprefix}{arXiv:}
\providecommand{\URLprefix}{URL: }
\providecommand{\Pubmedprefix}{pmid:}
\providecommand{\doi}[1]{\href{http://dx.doi.org/#1}{\path{#1}}}
\providecommand{\Pubmed}[1]{\href{pmid:#1}{\path{#1}}}
\providecommand{\bibinfo}[2]{#2}
\ifx\xfnm\relax \def\xfnm[#1]{\unskip,\space#1}\fi
\bibitem[{Wollschlaeger et~al.(2017)Wollschlaeger, Sauter, and
  Jasperneite}]{Wollschlaeger2017}
\bibinfo{author}{M.~Wollschlaeger}, \bibinfo{author}{T.~Sauter},
  \bibinfo{author}{J.~Jasperneite},
\newblock \bibinfo{title}{{The Future of Industrial Communication: Automation
  Networks in the Era of the Internet of Things and Industry 4.0}},
\newblock \bibinfo{journal}{IEEE Industrial Electronics Magazine}
  \bibinfo{volume}{11} (\bibinfo{year}{2017}) \bibinfo{pages}{17--27}.
  \DOIprefix\doi{Mar}.
\bibitem[{Li et~al.(2021)Li, Tanghe, Suanet, Plets, Hoebeke, De~Poorter, and
  Joseph}]{Chenglong2021Reloc2}
\bibinfo{author}{C.~Li}, \bibinfo{author}{E.~Tanghe},
  \bibinfo{author}{P.~Suanet}, \bibinfo{author}{D.~Plets},
  \bibinfo{author}{J.~Hoebeke}, \bibinfo{author}{E.~De~Poorter},
  \bibinfo{author}{W.~Joseph},
\newblock \bibinfo{title}{{ReLoc 2.0: UHF-RFID Relative Localization for
  Drone-Based Inventory Management}},
\newblock \bibinfo{journal}{IEEE Transactions on Instrumentation and
  Measurement} \bibinfo{volume}{70} (\bibinfo{year}{2021})
  \bibinfo{pages}{1--13}.
\bibitem[{Zafari et~al.(2019)Zafari, Gkelias, and Leung}]{Zafari2019}
\bibinfo{author}{F.~Zafari}, \bibinfo{author}{A.~Gkelias},
  \bibinfo{author}{K.~K. Leung},
\newblock \bibinfo{title}{{A Survey of Indoor Localization Systems and
  Technologies}},
\newblock \bibinfo{journal}{IEEE Communications Surveys \& Tutorials}
  \bibinfo{volume}{21} (\bibinfo{year}{2019}) \bibinfo{pages}{2568--2599}.
\bibitem[{Li et~al.(2021)Li, Zhuang, Hu, Gao, Hu, Chen, He, Pei, Chen, Wang,
  Niu, Chen, Thompson, Ghannouchi, and El-Sheimy}]{You2021}
\bibinfo{author}{Y.~Li}, \bibinfo{author}{Y.~Zhuang}, \bibinfo{author}{X.~Hu},
  \bibinfo{author}{Z.~Gao}, \bibinfo{author}{J.~Hu}, \bibinfo{author}{L.~Chen},
  \bibinfo{author}{Z.~He}, \bibinfo{author}{L.~Pei}, \bibinfo{author}{K.~Chen},
  \bibinfo{author}{M.~Wang}, \bibinfo{author}{X.~Niu},
  \bibinfo{author}{R.~Chen}, \bibinfo{author}{J.~Thompson},
  \bibinfo{author}{F.~M. Ghannouchi}, \bibinfo{author}{N.~El-Sheimy},
\newblock \bibinfo{title}{{Toward Location-Enabled IoT (LE-IoT): IoT
  Positioning Techniques, Error Sources, and Error Mitigation}},
\newblock \bibinfo{journal}{IEEE Internet of Things Journal}
  \bibinfo{volume}{8} (\bibinfo{year}{2021}) \bibinfo{pages}{4035--4062}.
\bibitem[{De~Bast et~al.(2022)De~Bast, Vinogradov, and Pollin}]{DeBast2022}
\bibinfo{author}{S.~De~Bast}, \bibinfo{author}{E.~Vinogradov},
  \bibinfo{author}{S.~Pollin},
\newblock \bibinfo{title}{{Expert-Knowledge-Based Data-Driven Approach for
  Distributed Localization in Cell-Free Massive MIMO Networks}},
\newblock \bibinfo{journal}{IEEE Access} \bibinfo{volume}{10}
  (\bibinfo{year}{2022}) \bibinfo{pages}{56427--56439}.
\bibitem[{Li et~al.(2022)Li, De~Bast, Tanghe, Pollin, and
  Joseph}]{Chenglong2022}
\bibinfo{author}{C.~Li}, \bibinfo{author}{S.~De~Bast},
  \bibinfo{author}{E.~Tanghe}, \bibinfo{author}{S.~Pollin},
  \bibinfo{author}{W.~Joseph},
\newblock \bibinfo{title}{{Toward Fine-Grained Indoor Localization Based on
  Massive MIMO-OFDM System: Experiment and Analysis}},
\newblock \bibinfo{journal}{IEEE Sensors Journal} \bibinfo{volume}{22}
  (\bibinfo{year}{2022}) \bibinfo{pages}{5318--5328}.
\bibitem[{Miesen et~al.(2013)Miesen, Kirsch, and Vossiek}]{Miesen2013}
\bibinfo{author}{R.~Miesen}, \bibinfo{author}{F.~Kirsch},
  \bibinfo{author}{M.~Vossiek},
\newblock \bibinfo{title}{{UHF RFID Localization Based on Synthetic
  Apertures}},
\newblock \bibinfo{journal}{IEEE Transactions on Automation Science and
  Engineering} \bibinfo{volume}{10} (\bibinfo{year}{2013})
  \bibinfo{pages}{807--815}.
\bibitem[{Xiao et~al.(2012)Xiao, Wu, Yi, and Ni}]{Xiao2012}
\bibinfo{author}{J.~Xiao}, \bibinfo{author}{K.~Wu}, \bibinfo{author}{Y.~Yi},
  \bibinfo{author}{L.~M. Ni},
\newblock \bibinfo{title}{{FIFS: Fine-Grained Indoor Fingerprinting System}},
\newblock in: \bibinfo{booktitle}{2012 21st International Conference on
  Computer Communications and Networks (ICCCN)}, \bibinfo{year}{2012}, pp.
  \bibinfo{pages}{1--7}.
\bibitem[{Li et~al.(2022)Li, Tanghe, Fontaine, Martens, Romme, Singh,
  De~Poorter, and Joseph}]{Chenglong2022UWB}
\bibinfo{author}{C.~Li}, \bibinfo{author}{E.~Tanghe},
  \bibinfo{author}{J.~Fontaine}, \bibinfo{author}{L.~Martens},
  \bibinfo{author}{J.~Romme}, \bibinfo{author}{G.~Singh},
  \bibinfo{author}{E.~De~Poorter}, \bibinfo{author}{W.~Joseph},
\newblock \bibinfo{title}{{Multistatic UWB Radar-Based Passive Human Tracking
  Using COTS Devices}},
\newblock \bibinfo{journal}{IEEE Antennas and Wireless Propagation Letters}
  \bibinfo{volume}{21} (\bibinfo{year}{2022}) \bibinfo{pages}{695--699}.
\bibitem[{Altstidl et~al.(2021)Altstidl, Kram, Herrmann, Stahlke, Feigl, and
  Mutschler}]{Altstidl2021}
\bibinfo{author}{T.~Altstidl}, \bibinfo{author}{S.~Kram},
  \bibinfo{author}{O.~Herrmann}, \bibinfo{author}{M.~Stahlke},
  \bibinfo{author}{T.~Feigl}, \bibinfo{author}{C.~Mutschler},
\newblock \bibinfo{title}{Accuracy-aware compression of channel impulse
  responses using deep learning},
\newblock in: \bibinfo{booktitle}{2021 International Conference on Indoor
  Positioning and Indoor Navigation (IPIN)}, \bibinfo{year}{2021}, pp.
  \bibinfo{pages}{1--8}. \DOIprefix\doi{10.1109/IPIN51156.2021.9662545}.
\bibitem[{Kotaru et~al.(2015)Kotaru, Joshi, Bharadia, and Katti}]{Kotaru2015}
\bibinfo{author}{M.~Kotaru}, \bibinfo{author}{K.~Joshi},
  \bibinfo{author}{D.~Bharadia}, \bibinfo{author}{S.~Katti},
\newblock \bibinfo{title}{{SpotFi: Decimeter Level Localization Using WiFi}},
\newblock in: \bibinfo{booktitle}{Proceedings of the 2015 ACM Conference on
  Special Interest Group on Data Communication (SIGCOMM)},
  \bibinfo{year}{2015}, p. \bibinfo{pages}{269–282}.
\bibitem[{Garcia et~al.(2017)Garcia, Wymeersch, Larsson, Haimovich, and
  Coulon}]{Garcia2017}
\bibinfo{author}{N.~Garcia}, \bibinfo{author}{H.~Wymeersch},
  \bibinfo{author}{E.~G. Larsson}, \bibinfo{author}{A.~M. Haimovich},
  \bibinfo{author}{M.~Coulon},
\newblock \bibinfo{title}{{Direct Localization for Massive MIMO}},
\newblock \bibinfo{journal}{IEEE Transactions on Signal Processing}
  \bibinfo{volume}{65} (\bibinfo{year}{2017}) \bibinfo{pages}{2475--2487}.
  \DOIprefix\doi{10.1109/TSP.2017.2666779}.
\bibitem[{Li et~al.(2023)Li, De~Bast, Miao, Tanghe, Pollin, and
  Joseph}]{Chenglong2023}
\bibinfo{author}{C.~Li}, \bibinfo{author}{S.~De~Bast},
  \bibinfo{author}{Y.~Miao}, \bibinfo{author}{E.~Tanghe},
  \bibinfo{author}{S.~Pollin}, \bibinfo{author}{W.~Joseph},
\newblock \bibinfo{title}{{Contact-Free Multitarget Tracking Using Distributed
  Massive MIMO-OFDM Communication System: Prototype and Analysis}},
\newblock \bibinfo{journal}{IEEE Internet of Things Journal}
  \bibinfo{volume}{10} (\bibinfo{year}{2023}) \bibinfo{pages}{9220--9233}.
  \DOIprefix\doi{10.1109/JIOT.2023.3234041}.
\bibitem[{Wang et~al.(2016)Wang, Gao, and Mao}]{Wang2016}
\bibinfo{author}{X.~Wang}, \bibinfo{author}{L.~Gao}, \bibinfo{author}{S.~Mao},
\newblock \bibinfo{title}{{CSI Phase Fingerprinting for Indoor Localization
  With a Deep Learning Approach}},
\newblock \bibinfo{journal}{IEEE Internet of Things Journal}
  \bibinfo{volume}{3} (\bibinfo{year}{2016}) \bibinfo{pages}{1113--1123}.
\bibitem[{Dai et~al.(2022)Dai, He, Tran, Trigoni, and Markham}]{Dai2022}
\bibinfo{author}{Z.~Dai}, \bibinfo{author}{Y.~He}, \bibinfo{author}{V.~Tran},
  \bibinfo{author}{N.~Trigoni}, \bibinfo{author}{A.~Markham},
\newblock \bibinfo{title}{{DeepAoANet: Learning Angle of Arrival From Software
  Defined Radios With Deep Neural Networks}},
\newblock \bibinfo{journal}{IEEE Access} \bibinfo{volume}{10}
  (\bibinfo{year}{2022}) \bibinfo{pages}{3164--3176}.
\bibitem[{Wymeersch et~al.(2012)Wymeersch, Marano, Gifford, and
  Win}]{Wymeersch2012}
\bibinfo{author}{H.~Wymeersch}, \bibinfo{author}{S.~Marano},
  \bibinfo{author}{W.~M. Gifford}, \bibinfo{author}{M.~Z. Win},
\newblock \bibinfo{title}{{A Machine Learning Approach to Ranging Error
  Mitigation for UWB Localization}},
\newblock \bibinfo{journal}{IEEE Transactions on Communications}
  \bibinfo{volume}{60} (\bibinfo{year}{2012}) \bibinfo{pages}{1719--1728}.
  \DOIprefix\doi{Jun}.
\bibitem[{Wu et~al.(2021)Wu, Yi, Wang, You, Huang, Gao, and Liu}]{Chi2021}
\bibinfo{author}{C.~Wu}, \bibinfo{author}{X.~Yi}, \bibinfo{author}{W.~Wang},
  \bibinfo{author}{L.~You}, \bibinfo{author}{Q.~Huang},
  \bibinfo{author}{X.~Gao}, \bibinfo{author}{Q.~Liu},
\newblock \bibinfo{title}{{Learning to Localize: A 3D CNN Approach to User
  Positioning in Massive MIMO-OFDM Systems}},
\newblock \bibinfo{journal}{IEEE Transactions on Wireless Communications}
  \bibinfo{volume}{20} (\bibinfo{year}{2021}) \bibinfo{pages}{4556--4570}.
\bibitem[{Witrisal et~al.(2016)Witrisal, Meissner, Leitinger, Shen, Gustafson,
  Tufvesson, Haneda, Dardari, Molisch, Conti, and Win}]{Witrisal2016}
\bibinfo{author}{K.~Witrisal}, \bibinfo{author}{P.~Meissner},
  \bibinfo{author}{E.~Leitinger}, \bibinfo{author}{Y.~Shen},
  \bibinfo{author}{C.~Gustafson}, \bibinfo{author}{F.~Tufvesson},
  \bibinfo{author}{K.~Haneda}, \bibinfo{author}{D.~Dardari},
  \bibinfo{author}{A.~F. Molisch}, \bibinfo{author}{A.~Conti},
  \bibinfo{author}{M.~Z. Win},
\newblock \bibinfo{title}{{High-Accuracy Localization for Assisted Living: 5G
  systems will turn multipath channels from foe to friend}},
\newblock \bibinfo{journal}{IEEE Signal Processing Magazine}
  \bibinfo{volume}{33} (\bibinfo{year}{2016}) \bibinfo{pages}{59--70}.
\bibitem[{Zhang et~al.(2021)Zhang, Zheng, Qian, Zhang, Liu, Wu, and
  Yang}]{Widar3}
\bibinfo{author}{Y.~Zhang}, \bibinfo{author}{Y.~Zheng},
  \bibinfo{author}{K.~Qian}, \bibinfo{author}{G.~Zhang},
  \bibinfo{author}{Y.~Liu}, \bibinfo{author}{C.~Wu}, \bibinfo{author}{Z.~Yang},
\newblock \bibinfo{title}{{Widar3.0: Zero-Effort Cross-Domain Gesture
  Recognition with Wi-Fi}},
\newblock \bibinfo{journal}{IEEE Transactions on Pattern Analysis and Machine
  Intelligence}  (\bibinfo{year}{2021}) \bibinfo{pages}{1--18}.
\bibitem[{Qian et~al.(2018)Qian, Wu, Zhang, Zhang, Yang, and Liu}]{Qian2018}
\bibinfo{author}{K.~Qian}, \bibinfo{author}{C.~Wu}, \bibinfo{author}{Y.~Zhang},
  \bibinfo{author}{G.~Zhang}, \bibinfo{author}{Z.~Yang},
  \bibinfo{author}{Y.~Liu},
\newblock \bibinfo{title}{{Widar2.0: Passive Human Tracking with a Single Wi-Fi
  Link}},
\newblock in: \bibinfo{booktitle}{Proceedings of the 16th Annual International
  Conference on Mobile Systems, Applications, and Services (MobiSys)},
  \bibinfo{year}{2018}, p. \bibinfo{pages}{350–361}.
\bibitem[{Wu et~al.(2021)Wu, Zeng, Gao, Li, Li, Shah, Lu, and Zhang}]{Wu2021}
\bibinfo{author}{D.~Wu}, \bibinfo{author}{Y.~Zeng}, \bibinfo{author}{R.~Gao},
  \bibinfo{author}{S.~Li}, \bibinfo{author}{Y.~Li}, \bibinfo{author}{R.~C.
  Shah}, \bibinfo{author}{H.~Lu}, \bibinfo{author}{D.~Zhang},
\newblock \bibinfo{title}{{WiTraj: Robust Indoor Motion Tracking with WiFi
  Signals}},
\newblock \bibinfo{journal}{IEEE Transactions on Mobile Computing}
  (\bibinfo{year}{2021}) \bibinfo{pages}{1--17}.
\bibitem[{Sakhnini et~al.(2022)Sakhnini, De~Bast, Guenach, Bourdoux, Sahli, and
  Pollin}]{Sakhnini2022}
\bibinfo{author}{A.~Sakhnini}, \bibinfo{author}{S.~De~Bast},
  \bibinfo{author}{M.~Guenach}, \bibinfo{author}{A.~Bourdoux},
  \bibinfo{author}{H.~Sahli}, \bibinfo{author}{S.~Pollin},
\newblock \bibinfo{title}{{Near-Field Coherent Radar Sensing Using a Massive
  MIMO Communication Testbed}},
\newblock \bibinfo{journal}{IEEE Transactions on Wireless Communications}
  \bibinfo{volume}{21} (\bibinfo{year}{2022}) \bibinfo{pages}{6256--6270}.
\bibitem[{Conti et~al.(2021)Conti, Morselli, Liu, Bartoletti, Mazuelas,
  Lindsey, and Win}]{Conti2021}
\bibinfo{author}{A.~Conti}, \bibinfo{author}{F.~Morselli},
  \bibinfo{author}{Z.~Liu}, \bibinfo{author}{S.~Bartoletti},
  \bibinfo{author}{S.~Mazuelas}, \bibinfo{author}{W.~C. Lindsey},
  \bibinfo{author}{M.~Z. Win},
\newblock \bibinfo{title}{{Location Awareness in Beyond 5G Networks}},
\newblock \bibinfo{journal}{IEEE Communications Magazine} \bibinfo{volume}{59}
  (\bibinfo{year}{2021}) \bibinfo{pages}{22--27}.

\end{thebibliography}

\end{document}